\documentclass[sigconf]{acmart}
\usepackage{enumitem}
\usepackage{graphicx}
\usepackage{subcaption}
\usepackage{multirow}
\usepackage{placeins}
\usepackage{booktabs}
\usepackage{tabularx}
\usepackage{array}
\usepackage{multirow}
\usepackage{makecell}
\usepackage{siunitx}
\usepackage{adjustbox}
\usepackage{makecell}
\usepackage{balance}

\newcolumntype{Y}{>{\raggedright\arraybackslash}X}
\AtBeginDocument{%
  }

\copyrightyear{2026}
\acmYear{2026}
\setcopyright{cc}
\setcctype{by}
\acmConference[KDD '26]{Proceedings of the 32nd ACM SIGKDD Conference on Knowledge Discovery and Data Mining V.2}{August 09--13, 2026}{Jeju Island, Republic of Korea}
\acmBooktitle{Proceedings of the 32nd ACM SIGKDD Conference on Knowledge Discovery and Data Mining V.2 (KDD '26), August 09--13, 2026, Jeju Island, Republic of Korea}
\acmDOI{10.1145/3770855.3818078}
\acmISBN{979-8-4007-2259-2/2026/08}
\begin{document}

%%
%% The "title" command has an optional parameter,
%% allowing the author to define a "short title" to be used in page headers.
\title{AnchorSteer: Self-Discovered Concept Injection for Structure-Preserving Music Editing}

%%
%% The "author" command and its associated commands are used to define
%% the authors and their affiliations.
%% --- 2026-05-18: reordered each author block to (\author -> \authornote -> \orcid -> \affiliation -> \email) so acmart renders email at the bottom, matching ACM convention (e.g., KDD 2025 papers) ---
\author{Chih-Heng Chang}
\authornote{These authors contributed equally to this work.}
\orcid{0009-0007-2791-6771}
\affiliation{%
  % \department{Graduate Institute of Communication Engineering}
  \institution{National Taiwan University}
  \city{Taipei}
  \country{Taiwan}}
\email{r14942077@ntu.edu.tw}

\author{Keng-Seng Ho}
\authornotemark[1]
\orcid{0009-0003-0085-4663}
\affiliation{%
  % \department{Graduate Institute of Communication Engineering}
  \institution{National Taiwan University}
  \city{Taipei}
  \country{Taiwan}}
\email{r13942143@ntu.edu.tw}

\author{Chih-Yu Tsai}
\authornotemark[1]
\orcid{0009-0000-9524-2480}
\affiliation{%
  % \department{Graduate Institute of Communication Engineering}
  \institution{National Taiwan University}
  \city{Taipei}
  \country{Taiwan}}
\email{r13942126@ntu.edu.tw}

\author{Kuan-Lin Chen}
\orcid{0000-0002-0491-5713}
\affiliation{%
  % \department{Graduate Institute of Communication Engineering}
  \institution{National Taiwan University}
  \city{Taipei}
  \country{Taiwan}}
\email{r13942067@ntu.edu.tw}

\author{Yi-Hsuan Yang}
\orcid{0000-0002-2724-6161}
% --- Yi-Hsuan Yang affiliation per 2026-05-18 email from Eric: full institute name preferred when space allows ---
% \affiliation{%
%   \department{Department of Electrical Engineering}
%   \institution{National Taiwan University}
%   \city{Taipei}
%   \country{Taiwan}}
\affiliation{%
  % \department{Artificial Intelligence Center of Research Excellence}
  \institution{National Taiwan University}
  \city{Taipei}
  \country{Taiwan}}
\email{yhyangtw@ntu.edu.tw}

\author{Jian-Jiun Ding}
\authornote{Corresponding author.}
\orcid{0000-0003-4510-2273}
\affiliation{%
  % \department{Graduate Institute of Communication Engineering}
  \institution{National Taiwan University}
  \city{Taipei}
  \country{Taiwan}}
\email{jjding@ntu.edu.tw}

%%
%% Page header — concise author list.
\renewcommand{\shortauthors}{Chih-Heng Chang et al.}
%% No italics, no superscripts, not anonymous
%% Use footnote or author note to identify equal contribution and/or contact author info

%%
%% The abstract is a short summary of the work to be presented in the
%% article.
\begin{abstract}
    % --- 2026-05-18: project page URL updated from anonymous briche402.github.io to permanent personal page brianchen1120.github.io/project/anchorsteer/ ---
    %    ... and the project demo page is \url{https://briche402.github.io/}.
    Controllable music editing is to modify high-level attributes while strictly preserving rhythmic and melodic structures. However, this task is challenged by a semantic-structural entanglement: steering methods often degrade structure to achieve editing performance, while structural adaptors suppress semantic responsiveness. We propose \textbf{AnchorSteer}, a framework that disentangles this tension by coupling structural anchoring with self-discovered semantic steering. The proposed approach probes internal representations to extract interpretable, label-free concept vectors via a self-supervised reconstruction objective, isolating attributes without curated data. During editing, these portable, plug-and-play concept vectors are injected into diffusion hidden manifolds while a structural adaptor enforces consistency. Variants for unconditioned and conditioned injections are provided to balance robustness and semantic strength. Experiments on ZoME-Bench and subjective tests show that the proposed framework outperforms both steering-only and anchoring-only baselines, enabling significant semantic transformations with high-fidelity structural preservation. 
    % The source code and the model checkpoint will be provided in the final version and the project demo page is \url{https://brianchen1120.github.io/project/anchorsteer/}.

    %Controllable music editing requires modifying high-level musical attributes while strictly preserving structure such as rhythm and melody. In diffusion-based models, this objective is challenged by a fundamental trade-off: semantic steering methods often achieve attribute changes by altering temporal structure, while structure-anchoring methods preserve alignment but suppress semantic editability. We propose \textbf{AnchorSteer}, a framework that explicitly resolves this tension by coupling structure anchoring with self-discovered semantic steering. Our approach first learns interpretable semantic concept vectors through a self-supervised, label-free reconstruction objective, isolating attribute-specific information without curated data. During editing, these plug-and-play concept vectors are injected into the diffusion hidden states while a lightweight structural adaptor enforces temporal and melodic consistency, with both unconditioned and conditioned injection variants that trade off structural robustness and semantic strength. Experiments on ZoME-Bench and subjective evaluations show that AnchorSteer outperforms both steering-only and anchoring-only baselines, enabling strong semantic editing while maintaining high-fidelity structural preservation.
\end{abstract}

\ccsdesc[500]{Computing methodologies~Knowledge representation and reasoning}
\ccsdesc[500]{Computing methodologies~Semantic networks}

%%
%% Keywords. The author(s) should pick words that accurately describe
%% the work being presented. Separate the keywords with commas.
\keywords{Music editing; Diffusion models; Concept discovery; Structure preservation; Latent space steering; Self-supervised learning}

% \received{9 February 2026}
% \received[revised]{12 March 2026}
% \received[accepted]{5 June 2026}

%%
%% This command processes the author and affiliation and title
%% information and builds the first part of the formatted document.
\maketitle

% --- 2026-05-18: Resource Availability block per KDD 2026 Artifact Badging policy. Macro template taken verbatim from https://kdd2026.kdd.org/call-for-artifact-badging/. Must sit after \maketitle and before \section{Introduction}. Earlier placement (after Conclusion as \section*{Resource availability}) was incorrect and is commented out near the end of this file. ---
\newcommand\kddavailabilityurl{https://doi.org/10.5281/zenodo.20368360}
\ifdefempty{\kddavailabilityurl}{}{
\begingroup\small\noindent\raggedright\textbf{Resource Availability:}\\
Source code: \url{https://github.com/hengtsune1024/AnchorSteer} (Zenodo archive: \url{\kddavailabilityurl}). Project pages: \url{https://brianchen1120.github.io/project/anchorsteer/}.
\endgroup
}

\section{Introduction}
\label{Introduction}
With the rapid progress of text-to-music (TTM) generation \cite{liu2024audioldm, copet2023simple}, the community has increasingly shifted the interest from unconstrained synthesis to \emph{controllable music editing}. In music editing, the goal is to refine a given source by modifying a target attribute (e.g., instrument or genre) while strictly preserving the source’s musical identity, including its rhythm, melodic contour, and temporal structure. Achieving these dual objectives, precise semantic control together with high-fidelity structural preservation, remains a central challenge for diffusion-based audio models.

Existing approaches struggle to balance these two objectives. Training-free editors such as MusicMagus \cite{zhang2024musicmagus} and Melodia \cite{yang2025melodia} perform editing primarily through text guidance but often require detailed source captions or precise hyperparameter tuning, making them fragile in practice. To strengthen structural preservation, MuseControlLite \cite{tsai2025musecontrollite} fine-tunes an adaptor to inject explicit structural conditions (e.g., melody), substantially improving content consistency over purely text-guided methods. However, such  structural anchoring  results in an over-constrained latent space, where the model prioritizes structural conditions at the expense of high-level semantic prompts, leading to weak or inconsistent attribute transfer. This exposes a key problem:  can we enforce  structural preservation without compromising semantic editability?

\begin{figure}[t]
\centering
\includegraphics[width=\linewidth]{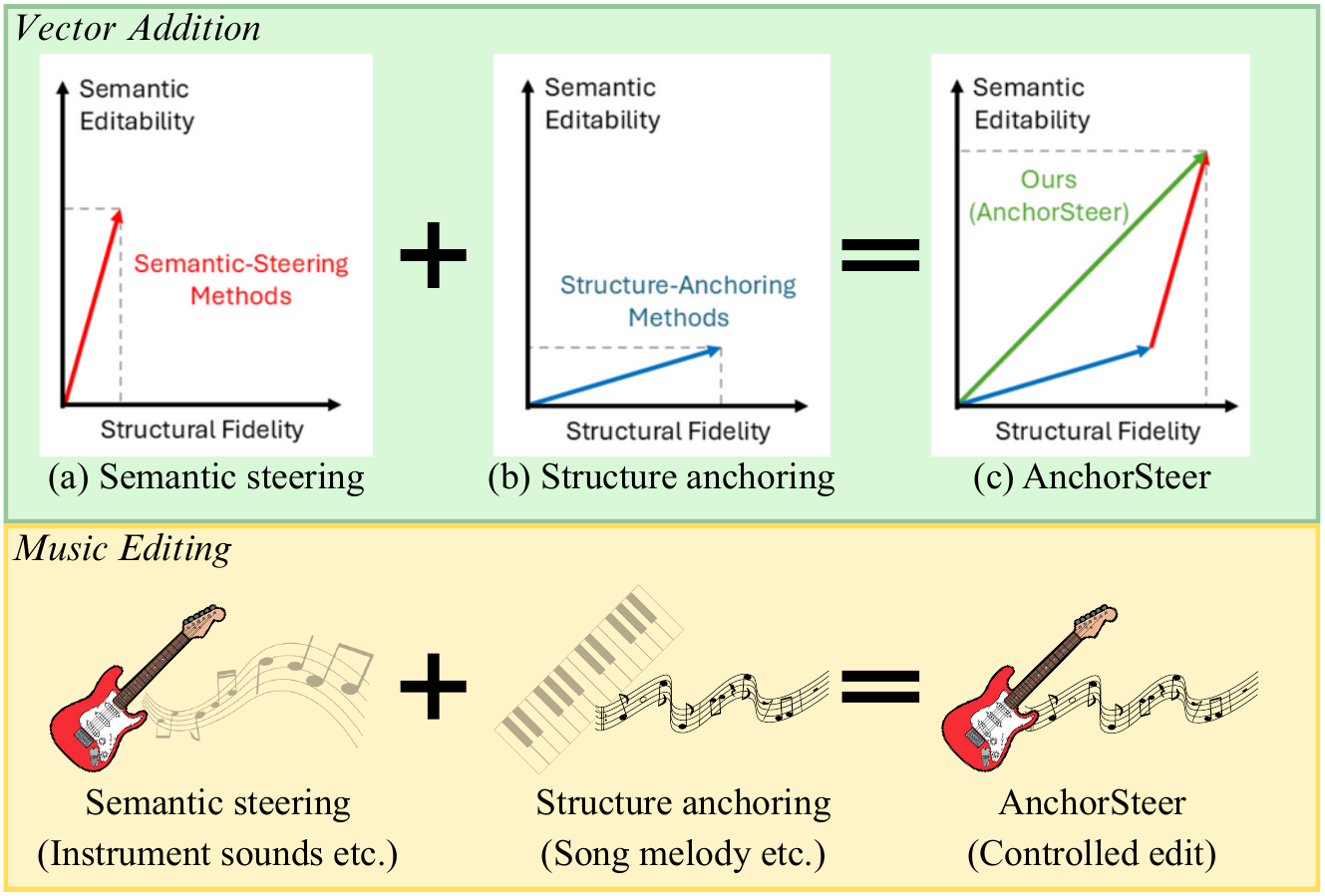}
\caption{Conceptual comparison of music editing paradigms.
  (a) Semantic steering (red) provides high editability but risks structural drift. (b) Structure anchoring (blue) preserves the musical scaffold but lacks semantic responsiveness due to over-constraints. (c) AnchorSteer (green) treats these as complementary forces. By steering atop an anchored base, the proposed method simultaneously achieves high editability and structural fidelity (upper-right corner).}
\Description{Three side-by-side plots illustrate the trade-off between Structural Fidelity (x-axis) and Semantic Editability (y-axis).
  (a) Semantic-steering methods are represented by a steep red vector starting from the origin, indicating high editability but low structural fidelity.
  (b) Structure-anchoring methods are represented by a shallow blue vector, indicating high structural fidelity but limited editability.
  (c) AnchorSteer, the proposed method, is represented by a green vector reaching the upper-right corner. It is visually depicted as the vector sum of the blue structure-anchoring vector and the red semantic-steering vector,
demonstrating that the proposed method achieves both high fidelity and high editability simultaneously.}
\label{fig:intro}
\end{figure}

\begin{figure}[t]
  \centering
  \includegraphics[width=\linewidth]{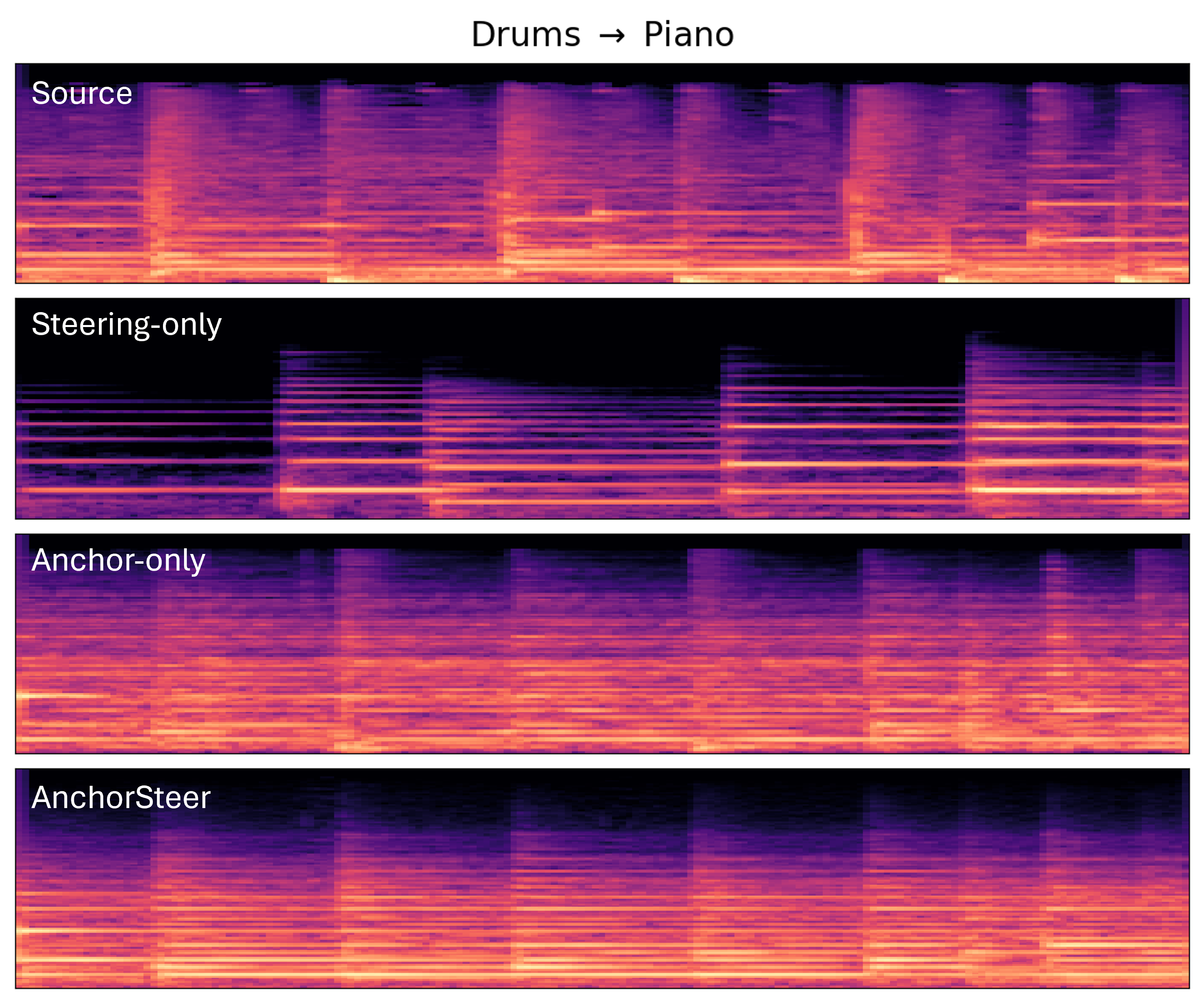}
  \caption{\textbf{Spectrogram comparison for Drums-to-Piano editing.} Top-to-bottom rows are: \emph{Source}, \emph{Steering-only}, \emph{Anchor-only} (MuseControlLite), and \emph{AnchorSteer}, respectively The Steering-only approach introduces harmonics but disrupts temporal alignment, whereas the Anchor-only approach preserves onsets but limits semantic transfer. AnchorSteer maintains the temporal scaffold while successfully injecting piano-like harmonics.}
  \Description{Figure~2 contrasts the failure modes of steering-only and anchoring-only for Drums$\rightarrow$Piano editing. Steering-only produces clear harmonic patterns but drifts from the source temporal structure, whereas anchoring preserves onsets but limits semantic transfer; AnchorSteer mitigates both by adding piano-like harmonics while maintaining the anchored structure.}
  \label{fig:mel}
\end{figure}

In parallel, semantic steering has recently matured in computer vision \cite{haas2024discovering, kwon2022diffusion, li2024self}: self-discovered directions or concept modules can be extracted from a pre-trained diffusion model using only the generations of the model, and then injected at inference time, without curated datasets, to induce strong high-level changes. Despite its success in vision, this paradigm has not been systematically validated for music editing, where the requirement is editing rather than unconstrained attribute-aligned synthesis. We therefore investigate \textit{whether CV-style self-discovered steering} \cite{li2024self} \textit{can be directly transferred to music}. Our findings suggest it \emph{cannot}: Steering alone often degenerates into re-synthesis, because musical semantics (e.g., instrument identity) are tightly coupled with temporal events; an unconstrained semantic force can satisfy the target attribute by rewriting the underlying temporal structure instead of preserving the source. Figure \ref{fig:mel} visualizes this failure mode: steering introduces piano-like harmonic patterns but breaks alignment with the source temporal structure, indicating a loss of edit fidelity.

These observations reveal a fundamental trade-off summarized in Figure \ref{fig:intro}: Semantic steering enables strong attribute changes but risks drifting away from the source structure, whereas structure anchoring preserves temporal patterns but can become over-constrained and yield weak semantic transfer. To resolve this tension, we propose AnchorSteer, a framework that explicitly decomposes editing into a structural scaffold (\emph{anchoring}) and a complementary semantic force (\emph{steering}). Concretely, we anchor generation with MuseControlLite while injecting self-discovered concept modules into diffusion hidden states to supply semantic steering. We further provide a portable ``plug-and-play'' injection module with two operating points: An \emph{unconditioned} injection that favors structural fidelity under strong anchoring, and a \emph{conditioned} injection that produces stronger and more robust semantic transfer by adapting the semantic force to the current hidden state, making it effective precisely in the over-constrained regime Figure \ref{fig:mel}. Overall, AnchorSteer offers, to our knowledge, a novel framework that makes self-discovered semantic steering practically usable for diffusion-based music editing by coupling it with explicit structural anchoring and state-dependent injection.

% --- P1 (qSR1 W1): added framing paragraph to emphasize knowledge discovery / semantic representation perspective ---
% --- P1 revision (R8 rebuttal-scope): trimmed expansions beyond rebuttal; previous version commented out below for diff ---
% While our experiments target music editing, the technical kernel of this work lies in the \emph{discovery, representation, and utilization} of interpretable, label-free concept vectors from a pretrained text-to-music diffusion backbone. These concept vectors are extracted via a self-supervised reconstruction objective without any curated labels, and serve as portable, plug-and-play semantic representations that can be injected into the model's hidden states. AnchorSteer thus connects to a broader line of research on extracting structured high-level semantics from large pretrained generative models---a setting where the additional requirement of structural preservation poses a distinctive challenge.
While our experiments target music editing, the technical kernel of this work lies in the \emph{discovery, representation, and utilization} of interpretable, label-free concept vectors from a pretrained text-to-music diffusion backbone. These concept vectors are extracted via a self-supervised reconstruction objective without any curated labels, and serve as portable, plug-and-play semantic representations that can be injected into the model's hidden states to enable controllable generation. From this perspective, the work is centered on the discovery, representation, and utilization of high-level concepts.

Our main contributions are summarized as follows.
% --- P3 (1QH1 W1): rewrote contributions to explicitly list hidden-state editing direction discovery as an independent contribution; original 3 bullets commented out below for diff ---
%\begin{itemize}[nosep]
%  \item  We propose AnchorSteer, a framework that couples structural anchoring with self-discovered semantic steering for diffusion-based music editing.
%  \item  We propose a plug-and-play injection module with two operating points: Unconditioned injection prioritizes structural fidelity, while conditioned injection enhances semantic transfer under structural constraints.
%  \item Comprehensive evaluation on ZoME-Bench demonstrating state-of-the-art semantic editing with practical structural preservation, validated by both objective metrics and subjective evaluation.
%\end{itemize}
\begin{itemize}[nosep]
  \item We identify reusable, label-free editing directions in the hidden states of text-to-music diffusion models, and propose a self-supervised reconstruction objective that extracts them as interpretable concept vectors usable as portable semantic representations.
  \item We propose \textbf{AnchorSteer}, a structure-aware editing framework that couples self-discovered semantic steering with explicit structural anchoring, directly addressing the semantic-structural trade-off identified above.
  \item We design a plug-and-play injection module with two operating points: \emph{unconditioned} injection for structural fidelity and \emph{conditioned} injection for stronger semantic transfer under heavy anchoring.
  \item We demonstrate state-of-the-art semantic editing with practical structural preservation on ZoME-Bench, validated by both objective metrics and a subjective listening test with 28 participants.
\end{itemize}

\section{Related Work}
\label{Related Work}

\textbf{Text-to-Music Generation.}
%\label{Text-to-Music Generation}
%
Modern TTM models mainly follow two paradigms: Autoregressive generation over discrete audio tokens and diffusion-based generation via iterative denoising in a latent space. MusicGen \cite{copet2023simple} formulates TTM as Transformer-based token prediction on EnCodec codes \cite{defossez2022high} with codebook interleaving for efficient generation. In contrast, diffusion-based methods \cite{ho2020denoising}  generate in a compressed latent space via iterative denoising. Building on this backbone, AudioLDM2 \cite{liu2024audioldm} adopts latent diffusion with a UNet-style denoiser \cite{ronneberger2015u} and leverage self-supervised pretraining with AudioMAE-derived representations \cite{huang2022masked} to support multimodal conditioning. Stable Audio Open (SAO) \cite{evans2025stable} further provides an open TTM diffusion model trained on Creative Commons audio, which combines waveform VAE compression \cite{kingma2013auto}, a T5-based text encoder \cite{raffel2020exploring}, and a diffusion Transformer (DiT) denoiser \cite{peebles2023scalable}.

\textbf{Music Editing and Controllability.}
%\label{Music Editing and Controllability}
Controllable music editing can be broadly organized into three approaches. %: adaptor-based conditioning, attention-based sampling control, and denoising-process-based editing. 
First, \textbf{adaptor-based methods} introduce lightweight modules to inject edit-related cues with minimal tuning. Representative examples include AP-Adapter \cite{tsai2024audio} and ControlNet-inspired music conditioners for time-varying controls \cite{wu2024music, zhang2023adding}. MuseControlLite \cite{tsai2025musecontrollite} further builds lightweight conditioners on SAO. It uses RoPE (rotary positional embeddings) aware decoupled cross-attention \cite{su2024roformer} to incorporate explicit structural cues (e.g., melody and rhythm) and achieve strong structural preservation. 
Second, \textbf{attention-based sampling control} is training-free and steers editing by modifying attention behaviors during inference, e.g., enforcing cross-attention consistency for semantic steering (MusicMagus \cite{zhang2024musicmagus}) or selectively modulating self-attention with an attention repository to preserve temporal structure under edits (Melodia \cite{yang2025melodia}). Finally, \textbf{denoising-process-based editing} also operates at inference time but focuses on aligning the edit to the diffusion trajectory itself, typically via inversion or partial noising: DDPM-Friendly audio editing \cite{manor2024zero} maps the input to a noise/trajectory representation and then re-denoises under a new condition, while SDEdit \cite{meng2021sdedit} starts denoising from an intermediate noise level.

\textbf{Semantic Concept / Direction Discovery.}
\label{Semantic Concept / Direction Discovery}
Semantic concept discovery for controllable generation has long been explored in latent-variable settings (e.g., VAE-style representations \cite{kingma2013auto}), where attributes are encouraged to align with latent factors for traversal-based control. However, prior work stresses that disentanglement does not necessarily imply controllability \cite{pati2021disentanglement}. In music, EC$^2$-VAE \cite{yang2019deep} explicitly decompose latent variables into semantically constrained components (e.g., rhythm) to support more interpretable manipulation. More recently, diffusion models have shown promising structure in their hidden-state latent spaces, especially in computer vision: Haas et al. \cite{haas2024discovering} studied interpretable directions in diffusion semantic latent spaces (h-space) via analyses such as PCA, Jacobian-based singular vectors, or simple supervised differencing, while Kwon et al. \cite{kwon2022diffusion} proposed an asymmetric reverse process to expose semantic latent information. Li et al. \cite{li2024self} further presented a self-discovery framework that learns interpretable diffusion latent directions and injects them at inference to steer generation. Our work builds on these diffusion-direction insights and adapts them to music editing, emphasizing the coupling of latent semantic steering with structure-preserving lightweight conditioners to alleviate the editability-fidelity trade-off.
% --- P4 (1QH1 W2): added sentences relating Pianist Transformer to our work via self-supervision contrast ---
Relatedly, Pianist Transformer \cite{you2025pianist} also places self-supervision at the core of its system design, although in the form of large-scale pre-training. Both directions share the use of self-supervision as a core component, while its role differs here: we use a self-supervised reconstruction objective to identify concept vectors in the hidden space of a pretrained text-to-music model.

\section{Preliminary}
\label{Preliminary}

\textbf{Latent Diffusion Models} (LDMs)
\label{Latent Diffusion Models}
%Latent Diffusion Models (LDMs) 
operate in a compressed latent space to reduce computational complexity while maintaining high generation quality \cite{rombach2022high}. The training process consists of two stages: A forward process that progressively corrupts a clean latent sample $z_0$ compressed from data $x_0$ via an encoder \cite{esser2021taming, van2017neural} into Gaussian noise $z_t$ over $t$ timesteps, and a reverse process to reverse this corruption. The model is typically trained to predict the noise $\epsilon$ added at each step by minimizing a mean squared error loss \cite{ho2020denoising}:
\begin{equation}
\mathcal{L}=E_{z_0,t,c,\epsilon\sim N\left(\mathbf{0},\mathbf{I}\right)}\|\epsilon-\epsilon_\theta (z_t,t,c)\|^2,
  \label{eq:diffusion-loss}
\end{equation}
where $\epsilon_\theta$ denotes the denoising network and $c$ represents the conditioning inputs.
In the context of music generation, models often employ a DiT architecture \cite{peebles2023scalable, evans2025stable} to capture long-range dependencies. During inference, the model starts with random noise $z_T$ and iteratively refines it to reconstruct a coherent musical sequence. Specifically, text conditioning is usually handled via cross-attention layers \cite{vaswani2017attention}, where the model applies text embeddings to guide the generation.

\textbf{MuseControlLite}
%\label{MuseControlLite}
%MuseControlLite 
\cite{tsai2025musecontrollite} is a parameter-efficient fine-tuning mechanism designed to inject precise and time-varying controls, such as melody and rhythm, into diffusion models. Unlike heavy retraining approaches, it utilizes decoupled cross-attention adaptors. Explicit attributes are first extracted from reference audio (e.g., via the CQT for melody or beat detection for rhythm) and projected to form a condition sequence $C_{struct}=\left\{c_n\right\}_{n=1}^N$.
A critical innovation is the integration of RoPE \cite{su2024roformer} to strictly enforce temporal alignment. Positional information is explicitly encoded into the query $q$, key $k$, and value $v$ vectors via a time-dependent rotation matrix $R_\Theta$:
\begin{equation}
  q_m = R_{\Theta,m} W_q x_m, \quad k_n = R_{\Theta,n} W'_k c_n, \quad v_n = R_{\Theta,n} W'_v c_n \,,
  \label{eq:rope_projection}
\end{equation}
where $x_m$ denotes the intermediate feature of the backbone model, $W_q$ is the frozen pretrained projection, and $W^{\prime}_{k},W^{\prime}_{v}$ are the trainable adaptor projections. By encoding absolute and relative positions, $R_\Theta$ compels the attention map to align perfectly with the condition's temporal grid. Finally, this structural guidance is injected into the original text-conditioned flow $x_{text}$ via a zero-initialized convolution layer \cite{zhang2023adding} $Z_{\mathrm{CNN}}$:
\begin{equation}
x_{\mathrm{out}}=x_{\mathrm{text}}+Z_{\mathrm{CNN}}\left(\mathrm{Attention}\left(Q_{rope},K_{rope},V_{rope}\right)\right),
  \label{eq:musecontrollite}
\end{equation}
where $Q_{rope}=\left\{q_m\right\}_{m=1}^M,\ K_{rope}=\left\{k_n\right\}_{n=1}^N,\ V_{rope}=\left\{v_n\right\}_{n=1}^N$, and $M$ represents the sequence length of the generated musical audio features. This mechanism forces the attention map to strictly adhere to the temporal grid of $C_{struct}$, effectively \emph{locking} the musical structure for our subsequent semantic editing.
\begin{figure*}[t]
  \centering
  \includegraphics[width=0.8\textwidth]{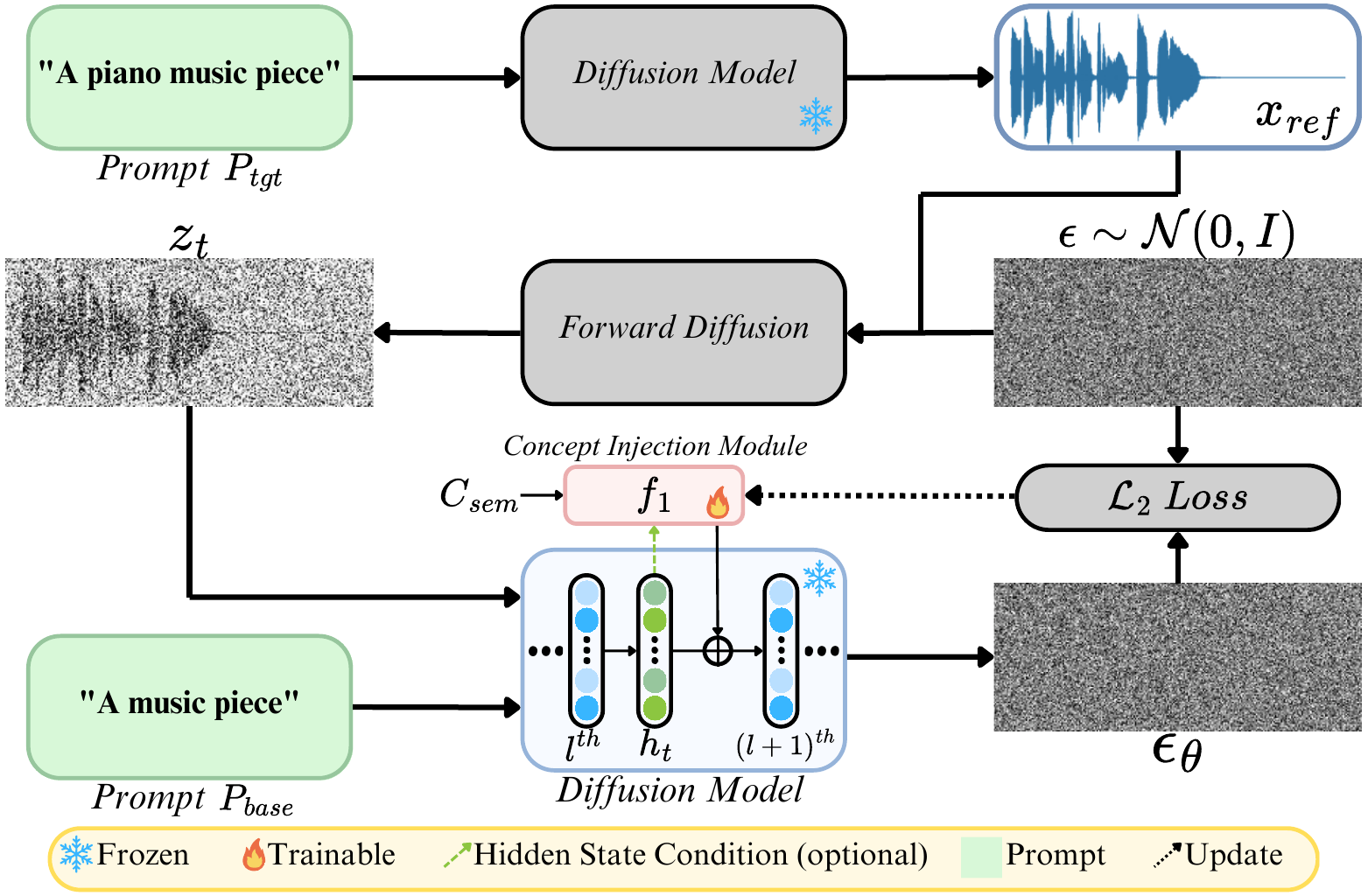}
  \caption{The overview of the proposed self-discovery approach to find concept vectors. We first generate a reference sample $x_{ref}$ using a specific target prompt $P_{tgt}$ (e.g., ``A piano music piece''). To capture the semantic attribute, we optimize a set of learnable concept injection modules $\mathcal{F}=\{f_l\}_{l\in\mathcal{L}}$ to reconstruct this reference from noise $z_t$, but conditioned on a generic base prompt $P_{base}$ (e.g., ``A music piece''). The diffusion model parameters are frozen, forcing $\mathcal{F}$ to learn the semantic difference between the generic and specific contexts by minimizing the reconstruction loss. The injection module can be conditioned on or independent of the current hidden states, depending on the selected parameterization strategy.}
  \Description{A flowchart showing the training pipeline: a reference music sample is generated from a target prompt, then a frozen diffusion model uses a generic base prompt and learnable injection modules to attempt reconstruction. The loss is calculated between the noise-corrupted reference and the model output.}
  \label{fig:self-discovery}
\end{figure*}
\section{Proposed Method}
\label{Proposed Method}
\subsection{Framework Overview}
\label{Framework Overview}
We propose AnchorSteer, a unified framework designed to resolve the inherent trade-off between semantic editability and structural fidelity by explicitly decomposing the generation process into two complementary forces. Our framework has two key components:
\begin{enumerate}
    \item A \textbf{Semantic-Steering} force, realized by our self-discovered concept injection modules, which drives the latent features toward the target attribute.
    \item A \textbf{Structure-Anchoring} backbone, realized by the MuseControlLite adaptor \cite{tsai2025musecontrollite}, which locks the temporal and melodic skeleton.
\end{enumerate}
Below, Section \ref{Self-Supervised Concept Vector Discovery} introduces a self-supervised approach to discover interpretable concept vectors, which serve as the Semantic-Steering force. After that, Section \ref{Synergistic Editing with Structural Constraints} presents the synergistic editing pipeline that couples this semantic steering with structure-anchoring to achieve robust attribute transfer.

\subsection{Concept Vector Discovery}
\label{Self-Supervised Concept Vector Discovery}

To construct our Semantic-Steering force, we present a self-supervised method to isolate high-level semantic attributes within the generative model's internal representation. Crucially, while the diffusion process operates in the compressed VAE latent space (z-space), our interventions target the model's deeper internal hidden-state space (h-space). Prior work \cite{haas2024discovering, li2024self} suggests that h-space encodes a richer semantic topology suitable for precise manipulation. The overall architecture is illustrated in Figure \ref{fig:self-discovery}.

\subsubsection{Contrastive Prompting and Reference Generation}

Let $C_{sem}$ denote a target semantic attribute (e.g., ``Piano''). We first construct a pair of contrastive text prompts: a target prompt $P_{tgt}$ that explicitly includes the attribute (e.g., ``A solo piano music piece''), and a base prompt $P_{base}$ that describes a generic context (e.g., ``A music piece'').

To establish a ground truth for the target attribute, we generate a set of reference samples $X_{ref}$ using the pretrained diffusion model $\epsilon_\theta$ conditioned on $P_{tgt}$. These samples serve as the training targets that embody the desired semantic feature.

\subsubsection{Concept Parameterization}
\label{Concept Parameterization}

We seek to identify a ``semantic gap'' in the model's hidden state space that corresponds to the transition from $P_{base}$ to $P_{tgt}$. To construct a potent Semantic-Steering module, we introduce a set of learnable concept injection modules $\mathcal{F}=\{f_l\}_{l\in\mathcal{L}}$, applied to a subset of layers $\mathcal{L}$ in the denoising model. These modules are designed to capture the semantic direction independent of the structural content. Let $h_l$ denote the intermediate hidden state of the $l$-th layer. We define the concept vector $\Delta h_l$ produced by the module $f_l$ as:
\begin{equation}
\Delta h_l=f_l(h_l).
  \label{eq:concept-vector}
\end{equation}
During the forward pass, these modules inject information additively into the intermediate feature maps. Formally, let $h_l\left(z_t,t,c\right)$ be the hidden state of the $l$-th layer at timestep $t$, conditioned on a general context $c$. The modified hidden state ${\hat{h}}_l$ is defined as:
\begin{equation}
{\hat{h}}_l=h_l\left(z_t,t,c\right)+\Delta h_l.
  \label{eq:inject}
\end{equation}
We explore two distinct parameterization strategies for the injection module $f_l$:
\begin{itemize}[nosep]
  \item  \textbf{Unconditioned Injection}: A strategy that utilizes a standalone vector $v_l$ that applies a static semantic bias independent of the current hidden state (i.e., $f_l\left(h_l\right)\equiv v_l$).
  \item  \textbf{Conditioned Injection}: A strategy that utilizes a lightweight network that dynamically computes the injection based on the input features (i.e., conditioned on $h_l$).
\end{itemize}
Detailed architectural specifications for both variants are provided in Section \ref{Implementation details}, while a comparative analysis of their effectiveness and trade-offs is presented in Section \ref{Comparison with State-of-the-Art}.

\subsubsection{Optimization Objective}

To learn these concept injection modules, we freeze the parameters of the diffusion model $\theta$ and optimize only the injection modules $\mathcal{F}$. We employ a reconstruction-based strategy where the model attempts to denoise the reference samples $X_{ref}$ (generated via $P_{tgt}$), but importantly, it is conditioned on the generic base prompt $P_{base}$.
The objective is to minimize the diffusion loss, forcing the concept injection modules to compensate for the semantic information missing from $P_{base}$ relative to $x\in X_{ref}$. The optimization problem is formulated as:
\begin{equation}
\mathcal{F}^\ast=\min_\mathcal{F}{E_{x\sim X_{ref},t,\epsilon}}\|\epsilon-\epsilon_\theta(z_t,t,P_{base},\mathcal{F})\|^2,
  \label{eq:concept-train}
\end{equation}
where $z_t$ is the noisy latent of a reference sample $x$ at timestep $t$, and $\epsilon$ is the ground truth noise.
By minimizing this reconstruction error, the optimized modules $\mathcal{F}^\ast$ effectively capture the semantic gap between the generic prompt and the specific attribute $C_{sem}$. Once optimized, these modules $\mathcal{F}^\ast$ serve as the standalone steering force for the subsequent inference stage.

This formulation offers three distinct advantages. First, it ensures ``plug-and-play'' universality: once trained, a concept module encapsulates a generalizable semantic feature that can be reused infinitely across diverse audio contexts without retraining. Second, it supports arbitrary concept discovery: our self-supervised approach can distill any user-defined attribute, ranging from specific instruments to abstract styles, solely through contrastive prompting, without requiring labeled data. Third, it eliminates manual tuning: unlike prior methods that often require heuristic selection of specific sensitive layers or timesteps, our framework learns a dedicated module for each layer that is effective across the entire diffusion process. This allows the mechanism to be applied uniformly without the need for manual layer or timestep selection.

\subsection{Structure-Anchored Steering}
\label{Synergistic Editing with Structural Constraints}
\begin{figure*}[t]
  \centering
  \includegraphics[width=0.7\textwidth]{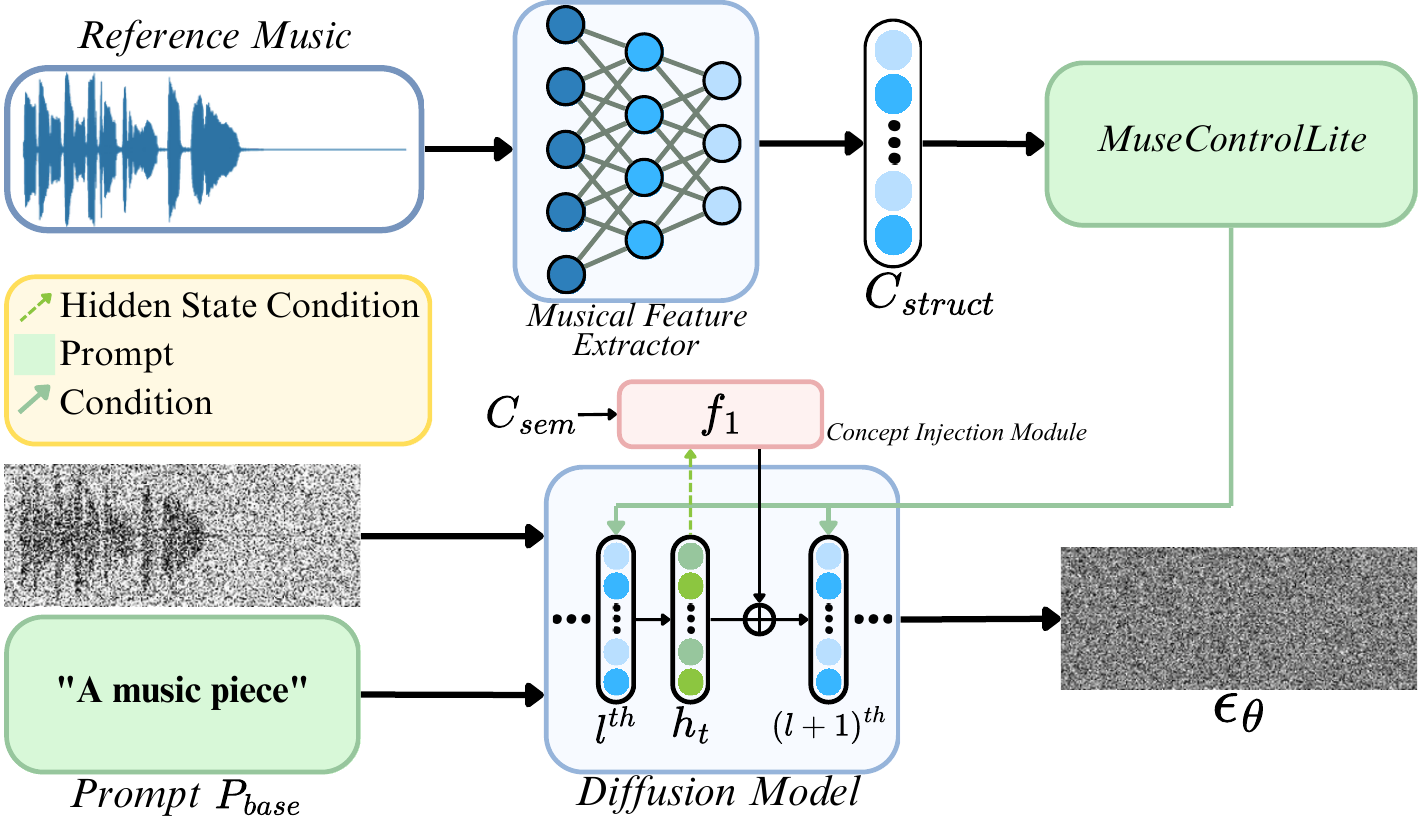}
  \caption{The overview of our synergistic editing pipeline. The framework integrates two guidance mechanisms to achieve controllable music editing. First, structural features (e.g., melody) are extracted from the reference music to form condition $C_{\text{struct}}$, which guides the MuseControlLite adaptor to enforce temporal consistency. Simultaneously, the pre-optimized concept injection module $f_l^*$ injects the concept vector into the hidden layers of the diffusion model.}
  \Description{A block diagram of the editing pipeline. It shows a reference music input being split into two paths: one for structural feature extraction leading to a MuseControlLite adaptor, and another through a pre-optimized concept injection module marked with an asterisk. Both paths converge on the hidden layers of a music diffusion model to produce the final edited output.}
  \label{fig:pipeline}
\end{figure*}
The core of our inference pipeline is the \textbf{synergistic coupling} of the two distinct mechanisms defined in Section \ref{Framework Overview}:
\begin{enumerate}
    \item \textbf{Structure-Anchoring Unit}: We utilize the MuseControlLite adaptor to serve as the structural backbone. By injecting explicit conditions $C_{struct}$ (e.g., melody), it enforces strict temporal alignment.
    \item \textbf{Semantic-Steering Unit}: Simultaneously, we activate the optimized concept injection modules $\mathcal{F}^\ast$ (Section \ref{Self-Supervised Concept Vector Discovery}) to act as the steering force that drives the attribute shift.
\end{enumerate}
By integrating these units, we achieve \textbf{structure-anchored steering}, allowing the model to traverse the semantic manifold while remaining tethered to the original musical skeleton.

During the denoising process, we first utilize the MuseControlLite adaptor to enforce structural fidelity. Explicit structural features are extracted from the source audio to form the condition sequence $C_{struct}$. These conditions are injected via cross-attention layers, producing a structurally guided hidden state representation.

Simultaneous to this structural conditioning, we inject the outputs of the optimized concept injection modules to induce the desired attribute shift. Let $\mathcal{F}^\ast=\{f_l^\ast\}_{l\in\mathcal{L}}$ denote the optimized modules obtained in the previous stage. For a target layer $l$, the final modified hidden state ${\hat{h}}_l$ is computed via additive injection:

\begin{equation}
\hat{h}_l=h_l(z_t,t,P_{edit},C_{struct}) + \lambda_{edit}f_l^*(h_l) \,,
  \label{eq:inference}
\end{equation}
where $h_l$ represents the intermediate hidden state. Compared to the general definition in Equation \ref{eq:inject}, the context $c$ is instantiated as the combination of structural inputs $C_{struct}$ (via MuseControlLite) and the text prompt $P_{edit}$. We use the base prompt $P_{base}$ as the default setting for all experiments. The term $f_l^\ast\left(h_l\right)$ represents the learned semantic modification, realized via either Conditioned or Unconditioned Injection (Section \ref{Concept Parameterization}). We set $\lambda_{edit}=1$, as the self-supervised reconstruction objective naturally learns the appropriate modification magnitude.

This integration effectively resolves the trade-off found in prior work. By injecting the concept vector directly into the h-space, we provide a potent steering force that reorients the generative trajectory along the desired semantic direction. %Crucially, 
While strong semantic shifts typically challenge structural integrity, the concurrent application of the structural adaptor acts as a robust anchor. It explicitly enforces temporal alignment, ensuring  the semantic transformation occurs within the boundaries of the original musical skeleton.

\section{Experimental Setup}
\label{Experimental Setup}

\subsection{Dataset}
\label{Dataset}

We evaluate our method on ZoME-Bench \cite{liu2024medic} under two editing tasks: instrument change and genre change. We first compute category frequencies on the original ZoME-Bench subsets (change-instrument and change-genre) and select the top-5 most frequent target categories for each task to ensure sufficient sample coverage. Since Stable Audio Open (SAO) produces audio with a fixed length of 47 seconds at 44.1 kHz, we then preprocess the selected audio tracks by splitting them into non-overlapping 47 second segments, treating each segment as one evaluation instance. Any remaining tail shorter than 47 seconds is discarded (i.e., no padding is applied). After segmentation, we obtain 578 segments for instrument change and 629 segments for genre change.

\subsection{Baselines and Comparison Paradigms}
\label{Baseline}

To comprehensively evaluate the performance of our framework and validate the trade-off hypothesis presented in Section \ref{Introduction}, we categorize the evaluated methods into three distinct editing paradigms:

\begin{itemize}
    \item \textbf{Structure-Anchoring Baseline}: Represented by MuseControlLite \cite{tsai2025musecontrollite}, which employs a structural adaptor to enforce strict temporal alignment but lacks latent steering mechanisms.
    \item \textbf{Semantic-Steering Baseline}: A configuration utilizing our Unconditioned Injection module \textit{without} the structural adaptor. This isolates the effect of pure latent steering (the red vector in Figure \ref{fig:intro}).
    \item \textbf{Synergistic Framework (Ours)}: Our full AnchorSteer method coupling the structural adaptor with concept injection. We evaluate both \textit{Unconditioned} and \textit{Conditioned} variants (details in Section \ref{Implementation details}).
\end{itemize}

Additionally, we compare our approach against established external baselines to ensure a comprehensive assessment: DDPM-Friendly \cite{manor2024zero}, SDEdit \cite{meng2021sdedit}, and MusicMagus \cite{zhang2024musicmagus}. We limit our comparison to methods with publicly available official implementations to ensure reproducibility. Consequently, some recent works (e.g., Melodia \cite{yang2025melodia}) are not included. We assess the performance of all methods using both objective metrics and subjective evaluations.

\subsection{Objective Evaluation Metric}
\label{Objective Evaluation metric}

To evaluate the edited audio, we adopt 3 commonly used metrics. CLAP \cite{wu2023large} measures semantic alignment between the audio and target attributes (higher is better). % --- P9 (6UJu Q1/Q4): specify exact LAION-CLAP checkpoint and reported zero-shot ESC-50 accuracy ---
Specifically, we use LAION-CLAP's \texttt{music\_audioset\_epoch\_15\_esc\_90.14.pt} checkpoint, which reports 90.14\% zero-shot accuracy on ESC-50 and is commonly adopted in music editing-related works for evaluating high-level attributes. LPAPS \cite{iashin2021taming, paissan2023audio} quantifies perceptual timbre differences between the original and edited audio to reflect audio quality and naturalness (lower is better). Chroma \cite{copet2023simple} assesses the preservation of harmonic content and pitch-class structure relevant to structural preservation (higher is better). 

Moreover, we introduce three CLAP-based advanced metrics, $\Delta$CLAP$_T$, $\Delta$CLAP$_S$, and GAP\cite{ezra2025freesliders}, to more precisely characterize the semantic shift induced by editing. Compared to absolute CLAP scores, which can be affected by how close the source audio is to the target concept, these metrics quantify (i) how much the edited audio improves alignment with the target attribute, (ii) how much it suppresses the original attribute, and (iii) the gap between the two. This provides a more comprehensive view of an editing method's effectiveness and capability.

Let $x_{src}$ denote the source audio and $x_{edit}$ the edited audio. Let $C_{tar}$ be the target attribute and $C_{src}$ the original attribute. We define:
\begin{align}
\Delta \mathrm{CLAP}_T &= \mathrm{CLAP}(x_{edit}, C_{tar}) - \mathrm{CLAP}(x_{src}, C_{tar}), \\
\Delta \mathrm{CLAP}_S &= \mathrm{CLAP}(x_{edit}, C_{src}) - \mathrm{CLAP}(x_{src}, C_{src}), \\
\mathrm{GAP} &= \Delta \mathrm{CLAP}_T - \Delta \mathrm{CLAP}_S .
\end{align}
Here, $\Delta$CLAP$_T$ measures the improvement in semantic alignment with the target attribute after editing (higher is better), reflecting the model's ability to steer the audio toward $C_{tar}$. $\Delta$CLAP$_S$ captures how the alignment to the original attribute changes after editing; ideally it should be small or negative. % --- P8 (6UJu Q3): added explicit interpretation of negative Delta CLAP_S ---
A negative $\Delta$CLAP$_S$ indicates that $\mathrm{CLAP}(x_{src}, C_{src}) > \mathrm{CLAP}(x_{edit}, C_{src})$, i.e., the original source attribute has been more successfully suppressed after editing. Finally, GAP jointly considers both moving toward the target and moving away from the original attribute. A larger GAP indicates that editing succeeds in increasing target alignment while reducing the influence of the original attribute, thus better reflecting overall editing capability.

\begin{table*}[t]
  \centering
  \caption{Validation of the Synergistic Design. We compare the three editing paradigms from Figure \ref{fig:intro}. We report GAP and $\Delta$CLAP metrics (defined in Section \ref{Objective Evaluation metric}) to measure the net attribute shift, alongside LPAPS and Chroma for structural fidelity. Bold indicates the best result, \underline{underline} the second best.}
  \Description{Quantitative comparison of three editing paradigms. The table highlights the trade-off: the Steering baseline has high GAP but poor structure (low Chroma); the Anchoring baseline has perfect structure (high Chroma) but poor GAP; the Synergistic framework achieves the best balance with high GAP and high Chroma.}
  \label{tab:module_combination}

  \small
  \setlength{\tabcolsep}{3pt}
  \renewcommand{\arraystretch}{1.05}

  \begin{tabular}{@{}l@{\hspace{4pt}}*{12}{c}@{}}
    \toprule
    & \multicolumn{6}{c}{Instrument} & \multicolumn{6}{c}{Genre} \\
    \cmidrule(lr){2-7} \cmidrule(lr){8-13}
    Methods
    & CLAP$\uparrow$
    & {$\Delta$CLAP$_T\uparrow$}
    & {$\Delta$CLAP$_S\downarrow$}
    & {GAP$\uparrow$}
    & LPAPS$\downarrow$
    & Chroma$\uparrow$
    & CLAP$\uparrow$
    & {$\Delta$CLAP$_T\uparrow$}
    & {$\Delta$CLAP$_S\downarrow$}
    & {GAP$\uparrow$}
    & LPAPS$\downarrow$
    & Chroma$\uparrow$ \\
    \midrule
    Steering baseline
      & \textbf{0.389} & \textbf{0.265} & \underline{0.001} & \textbf{0.263} & 12.213 & 0.091
      & \textbf{0.311} & \textbf{0.049} & \textbf{-0.040} & \textbf{0.089} & 11.671 & 0.098 \\
    Anchoring baseline
      & 0.250 & 0.126 & 0.013 & 0.113 & \textbf{9.828} & \textbf{0.488}
      & 0.293 & 0.032 & 0.000 & 0.032 & \textbf{9.607} & \textbf{0.467} \\
    \textbf{AnchorSteer}
      & \underline{0.320} & \underline{0.195} & \textbf{-0.003} & \underline{0.198} & \underline{10.346} & \underline{0.470}
      & \underline{0.301} & \underline{0.040} & \underline{-0.033} & \underline{0.073} & \underline{10.284} & \underline{0.406} \\
    \bottomrule
  \end{tabular}
\end{table*}

\begin{table*}[t]
  \caption{Objective comparison with State-of-the-Art on Instrument and Genre editing tasks. All metrics are detailed in Section \ref{Objective Evaluation metric}. We report semantic editability (CLAP, GAP$\uparrow$) and structural fidelity (LPAPS$\downarrow$, Chroma$\uparrow$). \textbf{Bold} indicates the best result, \underline{underline} the second best.}
  \label{tab:main_results}
  \Description{Comparison table of the proposed method against SDEdit, DDPM-friendly, MusicMagus, and MuseControlLite. Results show that the proposed Conditioned Injection variant consistently achieves the highest GAP and CLAP scores across Instrument and Genre tasks, outperforming baselines in semantic editing while maintaining competitive structural fidelity metrics.}
  \centering

  \small
  \setlength{\tabcolsep}{3pt}
  \renewcommand{\arraystretch}{1.05}

  \begin{tabular}{@{}l@{\hspace{4pt}}*{12}{c}@{}}
    \toprule
    & \multicolumn{6}{c}{Instrument} & \multicolumn{6}{c}{Genre} \\
    \cmidrule(lr){2-7} \cmidrule(lr){8-13}
    Methods
    & CLAP$\uparrow$
    & {$\Delta$CLAP$_T\uparrow$}
    & {$\Delta$CLAP$_S\downarrow$}
    & {GAP$\uparrow$}
    & LPAPS$\downarrow$
    & Chroma$\uparrow$
    & CLAP$\uparrow$
    & {$\Delta$CLAP$_T\uparrow$}
    & {$\Delta$CLAP$_S\downarrow$}
    & {GAP$\uparrow$}
    & LPAPS$\downarrow$
    & Chroma$\uparrow$ \\
    \midrule
    SDEdit
      & 0.260 & 0.135 & 0.011 & 0.124 & 10.525 & 0.213
      & 0.298 & 0.036 & -0.020 & 0.056 & 9.891  & 0.239 \\
    DDPM-friendly
      & 0.261 & 0.136 & 0.021 & 0.115 & \underline{9.127} & \underline{0.481}
      & 0.273 & 0.011 & -0.008 & 0.020 & \underline{8.279} & \textbf{0.561} \\
    MusicMagus
      & 0.217 & 0.092 & 0.046 & 0.047 & \textbf{7.774} & 0.395
      & 0.283 & 0.021 & \underline{-0.047} & 0.068 & \textbf{7.051} & 0.375 \\
    MuseControlLite
      & 0.250 & 0.126 & 0.013 & 0.113 & 9.828 & \textbf{0.488}
      & 0.293 & 0.032 & 0.000 & 0.032 & 9.607 & \underline{0.467} \\
    \midrule
    \makecell[l]{\textbf{Ours (Uncond.)}}
      & \underline{0.320} & \underline{0.195} & \underline{-0.003} & \underline{0.198} & 10.346 & 0.470
      & \underline{0.301} & \underline{0.040} & -0.033 & \underline{0.073} & 10.284 & 0.406 \\
    \makecell[l]{\textbf{Ours (Cond.)}}
      & \textbf{0.395} & \textbf{0.270} & \textbf{-0.008} & \textbf{0.279} & 11.852 & 0.238
      & \textbf{0.317} & \textbf{0.056} & \textbf{-0.081} & \textbf{0.136} & 10.992 & 0.217 \\
    \bottomrule
  \end{tabular}
\end{table*}

\subsection{Implementation details}
\label{Implementation details}

We employ the pre-trained SAO \cite{evans2025stable} as our backbone  model, which operates on a latent space compressed by a VAE. For structural guidance, we utilize the official implementation of MuseControlLite, employing its pre-trained adaptors to extract and inject melody, rhythm, and dynamics conditions. All experiments are implemented using PyTorch \cite{paszke2019pytorch} and the Hugging Face Diffusers library \cite{von-platen-etal-2022-diffusers}. Inference is performed using 50 denoising steps.

As mentioned in Section \ref{Concept Parameterization}, we implement two variants of the concept injection module:

\begin{itemize}
    \item Unconditioned Injection: This variant is implemented as a standalone learnable parameter $v_l\in R^{T\times D}$ matching the temporal (T) and channel (D) dimensions of the target hidden states. The vector is initialized to zeros and is broadcast across the batch dimension during the forward pass to be added to the hidden features.
    \item Conditioned Injection: This variant is implemented as a lightweight bottleneck Transformer. It comprises a down-projection layer (reducing input dimension to 256), a single Transformer encoder layer (8 attention heads, 512 feed-forward dimension), and an up-projection layer \cite{houlsby2019parameter}. Absolute positional embeddings are added to preserve temporal information, and the output is scaled by a learnable factor initialized to $10^{-4}$.
\end{itemize}
We inject the chosen module into all 24 hidden layers of the model.
% --- P7 (6UJu W1/W3): added prompt templates, 47s sample length, and informal quality verification; original sentence commented out below for diff ---
% For the self-supervised discovery stage, we generate 1,000 reference audio samples for each target concept using the base model.
For the self-supervised discovery stage, we generate 1,000 reference audio samples (each 47\,s long) for each target concept using the base model, with the prompt templates ``\texttt{a solo \{concept\} music piece}'' and ``\texttt{a typical \{concept\} music piece}''---for example, ``\texttt{a solo piano music piece}'' or ``\texttt{a typical jazz music piece}''. We perform an informal verification by manually listening to 30 samples per concept, which were found to be of reasonable quality and correctness overall. We optimize the concept injection module using the AdamW optimizer \cite{loshchilov2017decoupled} with a learning rate of $5\times{10}^{-4}$, a weight decay of $1\times{10}^{-4}$, and a cosine learning rate scheduler \cite{loshchilov2016sgdr} with 100 warmup steps. The training runs for 20 epochs with a batch size of 2 and gradient accumulation steps set to 2.
%Regarding computational cost, 
The data generation phase takes approximately 1 hour, while the optimization process requires about 30 minutes on a single NVIDIA RTX 4080 GPU.
\section{Results}
\label{Results}

\subsection{Efficacy of Synergistic Design}
\label{Efficacy of Synergistic Design}

To strictly validate the effectiveness of our proposed framework, we compare three distinct editing paradigms corresponding to the conceptual models illustrated in Figure \ref{fig:intro}. This experiment serves as the primary verification of our central hypothesis: that semantic editability and structural fidelity are often conflicting objectives, and a synergistic design is required to reconcile them. % --- P2 (qSR1 W2): explicitly label this experiment as a standalone ablation of the two main modules ---
This experiment also serves as our standalone ablation: by comparing structure-anchoring-only, semantic-steering-only, and the full AnchorSteer framework, we validate both the individual roles of these two modules and their complementarity. Table \ref{tab:module_combination} presents the quantitative comparison between the \textbf{Semantic-Steering baseline}, the \textbf{Structure-Anchoring baseline}, and our \textbf{Synergistic Framework}, as defined in Section \ref{Baseline}.

As shown in Table \ref{tab:module_combination}, the Semantic-Steering baseline (Row 1) yields the largest semantic shift, with a high GAP score. However, this high editability comes at a severe cost to structural fidelity, evidenced by the lowest Chroma score and highest LPAPS distance. Without a structural anchor, the concept vector becomes an unconstrained generative force. As discussed in Section \ref{Introduction}, steering alone often degenerates into \textit{re-synthesis}: the model generates a new track that matches the text but fails to preserve temporal alignment and harmonic identity.

In contrast, the Structure-Anchoring baseline (Row 2) preserves structure extremely well, achieving the best Chroma and LPAPS scores, but drastically limits editability. The GAP score drops to 0.113 for instrument editing and 0.032 for genre editing, indicating that the system is \textit{over-constrained}: the  structural guidance enforces the skeleton so strongly that it suppresses the semantic effect of the text prompt and blocks the intended attribute change.

Our method (Row 3) resolves this trade-off by coupling the two mechanisms. Integrating concept injection with the MuseControlLite backbone recovers much of the semantic controllability, raising GAP to 0.198 for instrument editing and 0.073 for genre editing, while preserving high structural fidelity. This empirically validates the AnchorSteer trajectory in Figure \ref{fig:intro}: the structural adaptor anchors the musical skeleton, while the concept injection steers timbre and texture within those boundaries. Thus, our synergistic design effectively decouples semantic manipulation from structural generation, enabling more robust editing than single-paradigm methods.

\subsection{Comparison with State-of-the-Art}
\label{Comparison with State-of-the-Art}
Following the setup in Section \ref{Baseline}, we evaluate our framework against the baselines (DDPM-Friendly \cite{manor2024zero}, SDEdit \cite{meng2021sdedit}, MuseControlLite \cite{tsai2025musecontrollite} and MusicMagus \cite{zhang2024musicmagus}).
\subsubsection{Objective evaluation}
\label{Objective evaluation}
Table \ref{tab:main_results} reports quantitative results for instrument and genre editing. Our framework, especially the Conditioned Injection variant, achieves the highest GAP across tasks, driven by high $\Delta \text{CLAP}_T$ and low $\Delta \text{CLAP}_S$. This shows that the concept injection mechanism effectively steers generation toward the target semantics while suppressing the source attribute.

For structural fidelity, the results reveal a clear trade-off. Baselines like MuseControlLite and MusicMagus show excellent structural retention but low GAP, indicating overly conservative edits. Strict adherence to the structural adaptor (as in MuseControlLite) over-constrains the model and limits the spectral changes needed for the target attribute.

Our method allows more structural deviation than these rigid baselines, which is necessary for strong semantic edits such as timbre transfer or genre shift, where spectral and harmonic content must change to match the new concept. The observed structural divergence in our model thus reflects the required signal transformation to satisfy the prompt. By augmenting the structural backbone with explicit semantic steering, our approach strikes a practical balance, enabling effective attribute transfer where conservative baselines fail to produce meaningful changes.

\subsubsection{Subjective evaluation}
\label{Subjective evaluation}

To strictly validate the perceptual performance of our framework, we conducted a subjective listening test involving 28 participants recruited from our social networks, of whom 40\% possess a background in music. We randomly selected 6 audio clips from the evaluation dataset as source samples, comprising 3 instrument editing tasks and 3 style transfer tasks. We generated edited versions using the same 6 methods employed in the objective evaluation (including our two proposed variants). The target attributes for editing were determined based on the corresponding ground truth editing labels in the dataset. Participants were presented with the source audio and the edited samples in a randomized order and were asked to rate each sample on a 5-point Likert scale \cite{likert1932technique} based on the following three criteria:
\begin{itemize}
    \item Target Attribute Match (T): To what extent does the audio embody the specific target attribute?
    \item Content Consistency (C): Compared to the original reference, is the audio recognizable as the same song?
    \item Audio Quality (Q): How would you rate the overall sound quality and naturalness of the audio?
\end{itemize}
Subjective results, summarized in Table \ref{tab:subjective_results}, indicate that our Conditioned concept injection method delivers the strongest perceptual editing performance. It achieves the premier ratings for Target Attribute Match, the most critical metric for evaluating editing success, as it directly measures whether the intended attribute transformation is perceptually achieved, demonstrating that our approach provides the most effective semantic steering. Additionally, this variant secures the highest Audio Quality scores, ensuring that the enhanced editability yields a natural listening experience.

Regarding Content Consistency, the results mirror the findings in the objective evaluation (Section \ref{Objective evaluation}), further illuminating the inherent trade-off between structural preservation and semantic editability. MuseControlLite, serving as our structural backbone, exhibits the most rigid adherence to the reference structure. While this validates the robustness of our underlying anchor, its lower editing scores imply that such strict preservation constrains the model from fully implementing the requested attribute. In contrast, our full method exhibits a moderate variation in consistency relative to its backbone. This shift likely reflects a necessary trade-off: the concept injection actively modulates the musical texture to align with the target attribute.

\begin{table}
  \centering
  \caption{Subjective evaluation results. We report Mean Opinion Scores (MOS) on a 5-point Likert scale (higher is better) across three criteria: Target Attribute Match (T), Content Consistency (C), and Audio Quality (Q). \textbf{Bold} indicates the best result, \underline{underline} the second best.}
  \Description{Subjective evaluation table comparing SDEdit, DDPM-friendly, MusicMagus, MuseControlLite, and Ours (Unconditioned/Conditioned). Metrics include Target Attribute Match (T), Content Consistency (C), and Audio Quality (Q). The results show that Ours (Conditioned) achieves the highest scores in Target Attribute Match (3.60) and Audio Quality (3.31), indicating superior editing capabilities. MuseControlLite scores highest in Content Consistency (3.85) but lower in Attribute Match, reflecting its over-constrained nature.}
  \label{tab:subjective_results}
  \begin{tabular}{lccc}
    \toprule
    Methods & T $\uparrow$ & C $\uparrow$ & Q $\uparrow$ \\
    \midrule
    SDEdit & 2.92 & 2.11 & 3.02 \\
    DDPM-friendly & 3.16 & 3.17 & \underline{3.26} \\
    MusicMagus & 2.92 & \underline{3.57} & 2.85 \\
    MuseControlLite & 3.03 & \textbf{3.85} & 2.83 \\
    \midrule
    \textbf{Ours (Unconditioned)} & \underline{3.18} & 3.45 & 2.60 \\
    \textbf{Ours (Conditioned)} & \textbf{3.60} & 2.94 & \textbf{3.31} \\
    \bottomrule
  \end{tabular}
\end{table}

\subsection{Influence of Text Prompting Strategies}
\label{Influence of Text Prompting Strategies}

As discussed in Section \ref{Synergistic Editing with Structural Constraints}, our framework allows for flexibility in the choice of the text prompt input during the editing stage. To investigate whether textual guidance is necessary to support the latent injection, we evaluate three distinct prompting strategies:

\begin{itemize}
    \item \textbf{Base Prompt}: A generic description containing no specific semantic information (e.g., ``A music piece'').
    \item \textbf{Target Prompt}: A description that explicitly includes the desired target attribute.
    \item \textbf{Original Prompt}: A description corresponding to the original source audio before editing.
\end{itemize}

The quantitative results are summarized in Table \ref{tab:prompt_ablation}. We observe distinct behaviors between the two parameterization strategies regarding their interaction with textual cues. For the Conditioned Injection, the method demonstrates remarkable robustness to prompt selection. The editing performance remains consistently high across all three strategies, with GAP scores hovering around $0.20$--$0.21$. Notably, even when conditioned on the Original Prompt, which explicitly describes the source attribute and contradicts the target injection, the model maintains a high GAP of $0.2$. This indicates that the adaptive network successfully overrides the conflicting textual guidance.
In contrast, the Unconditioned Injection is more sensitive to textual conflicts. Using the Original Prompt yields a lower GAP compared to the Base Prompt, suggesting that the static vector is less effective at overriding explicit source descriptions. Therefore, utilizing a neutral Base Prompt is recommended for this variant to ensure optimal attribute transfer.
% --- P2 (qSR1 W2): tie Tables 1 and 4 together as the standalone ablation suite supporting key design choices ---
Taken together with the comparison in Section~\ref{Efficacy of Synergistic Design} (Table~\ref{tab:module_combination}), this prompt-strategy analysis (Table~\ref{tab:prompt_ablation}) supports the effectiveness of AnchorSteer's key design choices: the individual roles of the structural anchor and the semantic injection, their complementarity, and the behavior of the injection module under different prompting conditions.

\begin{table}[t]
    \centering
    % --- P2 (qSR1 W2): prefix caption with "Ablation:" to make ablation framing explicit ---
    % \caption{Influence of text prompting strategies. We compare Unconditioned vs. Conditioned Injection across Base, Target, and Original prompts, using CLAP and GAP to measure robustness against conflicting textual cues.}
    \caption{Ablation: influence of text prompting strategies. We compare Unconditioned vs. Conditioned Injection across Base, Target, and Original prompts, using CLAP and GAP to measure robustness against conflicting textual cues.}
    \Description{Ablation study table comparing Unconditioned and Conditioned injection modules across three prompt types: Base (generic), Target (specific), and Original (source description). The table reports CLAP, $\Delta\text{CLAP}_S$, $\Delta\text{CLAP}_T$, and GAP. The Conditioned Injection module demonstrates high robustness, maintaining consistent GAP scores across all prompts. In contrast, the Unconditioned Injection module shows sensitivity to conflicting guidance, with performance degrading significantly when using the Original prompt.}
    \label{tab:prompt_ablation}
    \small
    \renewcommand{\arraystretch}{1.15}
    \setlength{\tabcolsep}{4pt}

    \begin{tabularx}{\columnwidth}{Y l S[table-format=1.3] S[table-format=-1.3] S[table-format=1.3] S[table-format=1.3]}
    \toprule
    % --- 2026-05-19: arrows on $\Delta$CLAP$_T$ and $\Delta$CLAP$_S$ were swapped relative to Table 1/2 and the metric definition (line 492-493). Fixed: $\Delta$CLAP$_T$ uses \uparrow (higher = better target alignment), $\Delta$CLAP$_S$ uses \downarrow (lower = better source suppression). ---
    Module & Prompt & {CLAP$\uparrow$} & {$\Delta$CLAP$_T\uparrow$} & {$\Delta$CLAP$_S\downarrow$} & {GAP$\uparrow$} \\
    \midrule
    \multirow{3}{=}{\makecell[l]{Unconditioned\\injection}}
      & Base     & 0.311 & 0.118 & -0.018 & 0.135 \\
      & Target   & 0.306 & 0.113 & 0.011 & 0.102 \\
      & Original & 0.252 & 0.059 & 0.046 & 0.013 \\
    \addlinespace[2pt]
    \midrule
    \multirow{3}{=}{\makecell[l]{Conditioned\\injection}}
      & Base     & 0.356 & 0.163 & -0.044 & 0.208 \\
      & Target   & 0.367 & 0.174 & -0.037 & 0.211 \\
      & Original & 0.367 & 0.174 & -0.026 & 0.200 \\
    \bottomrule
    \end{tabularx}
\end{table}

\section{Limitations}
\label{Limitations}
We outline a few directions for further development. Concept discovery is a one-time offline cost (approx. 30 minutes per attribute), with learned modules reused as plug-and-play components in subsequent editing; because editing acts on hidden states rather than retraining, the framework is compatible with stronger future text-to-music backbones. Our experiments focus on single-attribute instrument and genre editing, and have not systematically evaluated more complex instructions or joint multi-attribute editing; preliminary trials injecting two concept modules simultaneously did not show clear improvements, possibly involving concept interference, weight balancing, and conflict disentanglement. The 47-second editable length is inherited from the SAO backbone and can be extended with future long-form audio models; for long-form segment-wise editing, maintaining inter-segment continuity remains an open challenge. The fidelity of discovered concepts is tethered to the base model’s generation distribution, and the editability--structure trade-off, though mitigated by AnchorSteer, is not fully eliminated.

\section{Conclusion}
\label{Conclusion}
We studied the fundamental tension in diffusion-based music editing between semantic editability and structural fidelity. To reconcile these conflicting objectives, we proposed \textbf{AnchorSteer}, which couples a structure-anchoring backbone (MuseControlLite) with self-discovered, plug-and-play concept injection that provides an explicit semantic steering force in the model’s hidden states. We further introduced two operating points—unconditioned injection favoring structural preservation and conditioned injection enabling stronger, more robust semantic transfer under strict constraints. Experiments on ZoME-Bench validate the synergistic design: AnchorSteer substantially improves semantic transfer (GAP 0.279) while maintaining practical structural preservation, and subjective listening tests further confirm strong target-attribute match (MOS 3.60/5) with competitive perceptual quality. Future work will focus on achieving zero-shot generalization and mitigating reference bias.

% --- 2026-05-18: Resource Availability has been moved to its KDD-mandated position (after \maketitle, before \section{Introduction}); also shortened per Brian's preference. Old version commented out below for diff. ---
% \section*{Resource availability}
% We release the source code, model checkpoints, and the generated reference audio used for self-supervised concept discovery. The code repository is publicly available at \url{[CODE_REPO_URL_TODO]}, and an archival release is permanently deposited on Zenodo with concept DOI \url{https://doi.org/[ZENODO_DOI_TODO]}. Audio samples and additional results are available at the project demo page \url{https://brianchen1120.github.io/project/anchorsteer/}. The ZoME-Bench evaluation benchmark is publicly released by its original authors.

% --- P15: Acknowledgements. Yi-Hsuan Yang's funding info filled in per 2026-05-18 email from Eric. Jian-Jiun Ding's funding info and computing resources still pending. Previous placeholder version commented out below for diff. ---
% \begin{acks}
% We acknowledge the use of generative AI technologies solely for the purpose of language editing and grammatical refinement. The final content was reviewed and verified by the authors.
% \end{acks}
% \begin{acks}
% This work was supported in part by [GRANT_AGENCY_TODO] under grant [GRANT_NUMBER_TODO]. We thank [COMPUTING_RESOURCES_TODO] for computational resources, and the participants of our subjective listening test for their time and feedback. We acknowledge the use of generative AI technologies solely for the purpose of language editing and grammatical refinement. The final content was reviewed and verified by the authors.
% \end{acks}
\begin{acks}
% Yi-Hsuan Yang is supported by grants from Google Asia Pacific, the National Science and Technology Council of Taiwan (NSTC 114-2628-E-002-013-MY3), and the Ministry of Education (MOE) of Taiwan (for Taiwan Centers of Excellence). Jian-Jiun Ding was supported by the National Science and Technology Council of Taiwan (NSTC 114-2221-E-002 -126 -MY3. 
The work is supported by grants from Google Asia Pacific, the National Science and Technology Council of Taiwan (NSTC 114-2628-E-002-013-MY3 and NSTC 114-2221-E-002-126-MY3), and the Ministry of Education (MOE) of Taiwan (for Taiwan Centers of Excellence). We thank the participants of our subjective listening test for their time and feedback. % We acknowledge the use of generative AI technologies solely for the purpose of language editing and grammatical refinement. The final content was reviewed and verified by the authors.
\end{acks}

\bibliographystyle{ACM-Reference-Format}
\balance
\bibliography{sample-base-cameraready}

\end{document}